# Analysis of Student Behaviour in *Habitable Worlds* Using Continuous Representation Visualization

**Zachary A. Pardos**
Graduate School of Education & School of Information
University of California, Berkeley
zp@berkeley.edu

**Lev Horodyskyj**
School of Earth and Space Exploration
Arizona State University

**ABSTRACT**: We introduce a novel approach to visualizing temporal clickstream behaviour in the context of a degree-satisfying online course, *Habitable Worlds,* offered through Arizona State University. The current practice for visualizing behaviour within a digital learning environment has been to generate plots based on hand engineered or coded features using domain knowledge. While this approach has been effective in relating behaviour to known phenomena, features crafted from domain knowledge are not likely well suited to make unfamiliar phenomena salient and thus can preclude discovery. We introduce a methodology for organically surfacing behavioural regularities from clickstream data, conducting an expert in-the-loop hyperparameter search, and identifying anticipated as well as newly discovered patterns of behaviour. While these visualization techniques have been used before in the broader machine learning community to better understand neural networks and relationships between word vectors, we apply them to online behavioural learner data and go a step further; exploring the impact of the parameters of the model on producing tangible, non-trivial observations of behaviour that are suggestive of pedagogical improvement to the course designers and instructors. The methodology introduced in this paper led to an improved understanding of passing and non-passing student behaviour in the course and is widely applicable to other datasets of clickstream activity where investigators and stakeholders wish to organically surface principal patterns of behaviour.

**NOTES FOR PRACTICE**

- Continuous representation visualization can produce a high-level view of emergent student behavior online without the need for defining features or tagging
- Differential visualization of passing and non-passing student course behaviors can help identify deep and shallow learning strategies and provide instructors with essential information for modifying the curricula to discourage strategies associated with failure
- Involving instructors in the tuning of the visualization and model parameters produces analyses with a desirable mixture of expected and unexpected, but explainable, patterns
- Layering on additional data, such when students create a discussion post, further contextualizes insight into student learning strategies from visualizations

## INTRODUCTION

A highly-touted benefit of a completely online or otherwise digital course has been that a detailed record of the interactions of learners with course materials is kept that can then be mined for potential actionable insights. It is therefore no surprise that descriptive statistics of the interactions between learners and pedagogical materials were among the first type of information sought by online course instructors. Among their questions of curiosity were: How did students utilize the course resources? How did struggling students differ in their usage patterns compared to passing students? Where did students need help and what can the data provide that can help make improvements to future offerings? While prior work has made headway, these questions largely remain unanswered. The current behaviour visualization practice of viewing state space diagrams of student transitions from one resource to another can be unsatisfying as it is ultimately a descriptive statistic summarization of behaviour that can struggle to surface less anticipated patterns of behaviour that require pattern recognition over a longer context window. Classification models of varying complexity can learn similarly complex behavioural patterns, and while they can at times convey enough corroborating evidence to convince researchers and educators that their predictions are robust, they rarely provide enough novel and interpretable information to affect an instructor's existing understanding of their own course. Linear representation learning approaches strike a happy medium between simple descriptions and the obscurity of more complex statistical models. As opposed to support vector machines and the cadre of neural network approaches labeled as "deep learning," representation learning algorithms (e.g., skip-grams) are in the class of simple linear feed-forward neural network models. While they are trained by optimizing a predictive outcome, it is the structure of the data found in the model's learned parameters (i.e., the representation) that is the artifact of value. In this paper, we show that when used in combination with dimensionality reduction techniques, robust emergent patterns in the data can be surfaced with an order of nuance not attainable with existing descriptive approaches.

## BACKGROUND

The practice of training a machine-learned classifier in predictive learning analytics has involved the customary process of feature engineering, a step of transforming data from its original form into a set of more abstract, hand-crafted features leveraging the domain knowledge of the researcher. This is often a process of aggregation, normalization, or combination of multiple attributes to extract essential descriptors from the original unstructured or semi-structured data (Freitag, 2000). The motivation for this process is often two-fold: (i) to bring the prior knowledge of the researcher to bear on the problem and (ii) to transform the original data into a form that is syntactically compatible with the chosen classifier(s). This standard practice can be seen in learning analytics work predicting student affect (Baker et al., 2012), drop-out (Boyer, Veeramachaneni, 2015), and question correctness (Stamper & Pardos, 2016), among many other applications of data mining in education (Koedinger et al., 2015). The premise of this approach is that by funnelling data through the prism of the researcher's domain knowledge, the engineered set of features will be a better representation for the prediction task than the original untreated data. "Better," in this case, defined as leading to better model fit or predictive generalization. The intuition being that there are certain hand-engineered transformations of the original attributes that closely relate to what is being predicted but would be difficult for a classifier to learn from the original data in the process of training. This approach has been quite effective in terms of producing classifiers of reasonable accuracy. When working with smaller datasets, the feature-engineering process can be seen as a type of non-linear human specified functional transformation between data and an intermediate representation which is closer to the target of prediction than the original data. In scenarios with smaller data, bringing the researcher's generalizations to bear to create these representations often leads to more robust models than attempting to statistically learn them.

Visualization, too, can be seen as a manual feature engineering process whereby the researcher brings her prior domain knowledge and hypotheses to bear on the transformation of the data from its original form. The difference between this process in the context of visualization and machine classification is that in visualization, this feature representation is being presented to a human learner, rather than a machine classifier, to understand and reason about. In the area of visualizing behaviour in an online course environment, which is the case study of this paper, common engineered features have been descriptive statistics of MOOC certification (Breslow et al., 2013), dwell time by resource category (Seaton et al., 2013), and counts of click-stream actions in quantized windows over time combined with summaries of forum activity (Crossley et al., 2016). Other work has taken an approach of visualizing behaviour using conditional probabilities (e.g. Markov models), which express the most common transitions from one resource to another (Köck & Paramythis, 2011; Caprotti, 2017) or a sub-sequence of commonly occurring transitions called motifs (Davis, Chen, Hauff, & Houben, 2016). In the case of the descriptive approach, a bar chart or scatter plot is used with the X-axis typically being a categorical (e.g., types of resources) or timescale (e.g., week 1 through 10) and the Y-axis, the attribute being summed (e.g. dwell time or # of actions). In Xu et al. (2014), for example, the Y-axis of their three scatter plots was the learners' course grade, with the X-axis being quiz clicks, lecture page views, and discussion page views. An overview of the past art on visualizing behaviours within Massive Open Online Courses can be found in Emmons, Light, & Börner (2017). The transition frequency approach can be seen as an attempt to disaggregate the descriptive approach and study the relationships between individual resources or resource types. In this approach, a graph is visualized with nodes representing resources and edges representing transitions. The frequency of the transition can be expressed with the thickness or length of the edge. In the case of a report on the first four years of MITx MOOCs (Chuang & Ho, 2016), a prominent figure had each node representing a course and the size of the node depicting the number of enrolments.

In graphs, the angular orientation between two vertices with respect to a frame of reference is often irrelevant. Applications like graph-viz (Ellson et al., 2001) use variants on force-directed algorithms that optimize angles and orientations for visual appeal but do not necessary encode any additional information. The only constraints are the set of edges and vertices and, optionally, the length of those vertices. By visualizing machine-learned representations instead of hand-specified features, we hypothesize that patterns of greater novelty and significance can be revealed. Work has begun to bridge visualization with modeling, using regression with pre-hypothesized features and outcomes of the environment (Fratamico, Perez, & Roll, 2017; Park et al., 2017) and using singular-value decomposition to study relationships between assessment constructs, or epistemic elements, based on their coded co-occurrences in discourse (Shaffer & Ruis, 2017). Bergner, Shu, & von Davier (2014) discuss the remaining difficulties of visualizing variable length learner clickstream data and the inadequacies of current methods. The representation approach introduced in this paper provides a method to address these difficulties.

The practice of feature engineering is often an informationally lossy one. For one, because much of feature engineering is based on aggregation and summarization, and because features created by hand are limited to the scope of the researcher's intuition and domain knowledge. When an ample degree of domain knowledge is present, feature engineering can be an effective way of explicitly defining relevant relationships in the data that a statistical learning approach might struggle to identify. For domain areas in which there is little theory or expertise, such as many behavioural contexts, feature engineering may not be a viable option. In these areas, features may be engineered which effectively describe known phenomena but are ineffective at describing unfamiliar phenomena, thus inhibiting downstream discovery. The general paradigm of using connectionist models (i.e., neural networks) to generate these features is called representation learning (Bengio, Courville, & Vincent, 2013). There has been no field where the lossiness of feature engineering has been made more apparent than in computer vision. The dominant domain expert intuition for the task of classifying what object is featured in an image was to engineer a set of features which describe the edges present in an image and to then present this set of edge descriptions to a classifier for training. The accuracy results of this approach were eclipsed by Convolutional Neural Networks (Krizhevsky, Sutskever, & Hinton, 2012), which automatically learn rich feature representations of the image from the original pixel data. We hypothesize that by using representation learning applied to student behavioural data in an online course, as it has been applied to problem interaction sequences in math tutoring systems (Pardos & Dadu, 2017; Pardos, Farrar, Kolb, Peh, & Lee, 2018) and course enrolment data (Pardos & Nam, 2017), the most significant features of behaviour can be revealed where they may have never been detected if first filtered through the prism of one's domain knowledge. The visualization of the learned representations in our approach is a visualization of the patterns or regularities of student behaviour learned by a model which combs through course clickstream, a kind of data hardly scrutable to an instructor in untreated form.

## METHODOLOGY

This paper introduces a novel technical methodology involving representation learning and visualization as well as a qualitative methodology in which the visualization is tuned and interpreted in close collaboration with the course instructor.

The technical methodology began with a dataset containing the chronological sequences of interactions of students with an online for-credit course offered by Arizona State University. A representation learning model common in computational linguistics, called a skip-gram (Mikolov, Yih, & Zweig, 2013), was applied to this behavioural sequence. This is a novel application of the algorithm as it is customarily applied to sequences of words, not behaviours. We hypothesize that the analogy to language representation will hold. This process learns a high dimensionality vector representation of every course element interacted with by students. These vectors were reduced to two dimensions for visualization using a non-linear dimensionality reduction technique called t-Stochastic Neighbourhood Embedding (t-SNE), designed to visualize hidden layers in a neural network while retaining significant structure (Maaten & Hinton, 2008).

The tuning and interpretation methodology involved close communication with a subject matter expert, the co-creator and instructor of the course. Since different hyperparameter values of the representation learning method can produce dramatically different vectors, and consequently dramatically different reduced dimensionality visualizations, different values for vector size and window size were used to produce 21 different visualizations of course behaviour that the course instructor was asked to then rate for the amount of novel information they contained. Additional guidance was given that favourable visualizations might be ones that both contained behavioural patterns the instructor was confident existed, , a priori, but also contained patterns that were not anticipated but still plausible. The purpose of the visualization was to surface information not already known by the expert. The rationale was that if the visualization depicted behavioural patterns known to be true by the expert, other aspects of the visualization may also be true but not yet known. After identifying useful parameters for the base models, visualizations of student discussion posts and differences between passing and failing students were designed in concert with the instructor. An interactive d3-based visualization was developed simultaneously with this research to enable the subject matter expert to inspect the visualization by hovering the mouse cursor over plot points to reveal semantic meta-information about the element, such as the name of the question and which lesson it belonged to. A colouring feature was also made available whereby the data points could be coloured by lesson or any other categorical feature of the element. The features of the in-house created d3 visualization[1] were very similar to features offered by a commercial software package called Tableau but had the benefit of being easily linked via URL in shared web documents and is now open sourced to the community.

## Representation Learning with Skip-Grams

When applied to natural language, a skip-gram model will learn a vector representation of a word based on the many contexts in which it appears across a large corpus of text. The prediction objective of the model is to, given an input word, predict the probability of words appearing in context with the input word. The probability of all words in the vocabulary must sum to one and the error of the model is calculated based on the probability values of the context words. The vector representation of the word is the output of the single hidden layer network that comprises the model. When two words share similar contexts, they will likely have similar learned vector representations. Often, synonyms of words will have vectors located close to one another in this vector space, also referred to as an embedding. The novel intuition of our application of this to student course sequences is that instead of learning the structure of language by training on sequences of words, we are learning the structure of learner behaviour from sequences of page views. It was previously found that clickstream behaviours within MOOCs could be predicted using a Recurrent Neural Network (RNN) with 70% accuracy, compared to the 45% accuracy provided by the expected path through the course when following the existing course structure (Tang, Peterson, & Pardos, 2017). This work builds on the observation that patterns exist in learner clickstream behaviours. Instead of focusing on prediction, in this work we seek to scrutinize course component embeddings to qualitatively learn what these patterns are and their relevance to the pedagogical design of the course. While RNNs are useful for prediction, skip-gram models are better suited for interpretation as they are linear models which create representations embedded in a vector space, subject to arithmetic and scalar manipulation. These representations, in language, are evaluated based on the percentage of pre-enumerated semantic and syntactic relationships they encode (Mikolov, Chen, Corrado, & Dean, 2013). In our application, the subject matter expert's implicit knowledge of the course serves as the validation and interpretation of plausible relationships not previously known, a methodological step not yet undertaken in computational linguistics.

[1] https://github.com/CAHLR/d3-scatterplot

Formally, a skip-gram is a simple feed-forward, three-layer neural network with one input layer, one hidden layer, and one output layer (Figure 1). The input, in our context, is a one-hot representation of the course element and the output can be described as a multi-hot representation of the specified number of elements in context. The number of elements in context is two times the window size, which is a hyperparameter of the model. The objective of the model is to predict the elements in context given the input element. Since multiple elements are being predicted, the loss function, categorical cross-entropy, is calculated for each element in context. The number of weights trained in the model does not increase with the window size since the same weights are used to make predictions of every element in context. The second major hyperparameter is the size of the hidden layer, which ultimately is equivalent to the number of dimensions of the learned continuous course element vector. The continuous vector is the weights associated with the edges stemming from the one-hot position of the element to all the nodes in the hidden layer. In the case of natural language, the inputs are words in a vocabulary, processed by sweeping sequentially through the words in a large corpus, where the model objective is to predict the words in context given an input word. In our online course interactions context, the inputs are elements a student interacts with and the model sweeps across chronological sequences of these elements. In our particular dataset of a course created in the Smart Sparrow course platform, the logged actions are students' interactions with screens, or pages, in the course's lessons containing simulations, practice problems, and graded assessments. Specifically, the inputs are screen IDs and the outputs being predicted are the screen IDs before and after the current screen ID as accessed by students. No correctness information is used in this model, as it is the student's navigational behaviour that is the focus, not their performance.

In a skip-gram, the vector representation of an input screen is defined as: $v_{w_I} = W^T \delta(w_I)$

Where $W^T$ is the left side weight matrix in Figure 1, indexed by the one-hot of the input screen, $\delta(w_I)$.

A softmax layer, common to classification tasks, is used to produce a probability distribution over screens to predict screens in context: $p(w_O|w_I) = \frac{\exp(W'\delta(w_o)v_{w_I})}{\sum_{j=1}^{V}\exp(W'\delta(w_j)v_{w_I})}$

For a given output screen, $w_O$, in the vocabulary, its probability is the exponential normalization defined by the exponentiation of the input screen's vector, $v_{w_I}$, multiplied by the output screen's vector, $W'\delta(w_O)$, divided by the sum of all screens' exponentiation of their output vector multiplied by the input vector. An output vector is the multiplication of the right-side weight matrix, $W'$, with a one-hot of the output screen, $w_O$.

The cross-entropy loss (i.e., log loss) across all students' sequence of screens, which is backpropagated, is:

$$C = -\sum_{s \in S} \frac{1}{T} \sum_{t=1}^{T} \sum_{-c \leq i \leq c, i \neq 0} \log p(w_{t+j}|w_t)$$

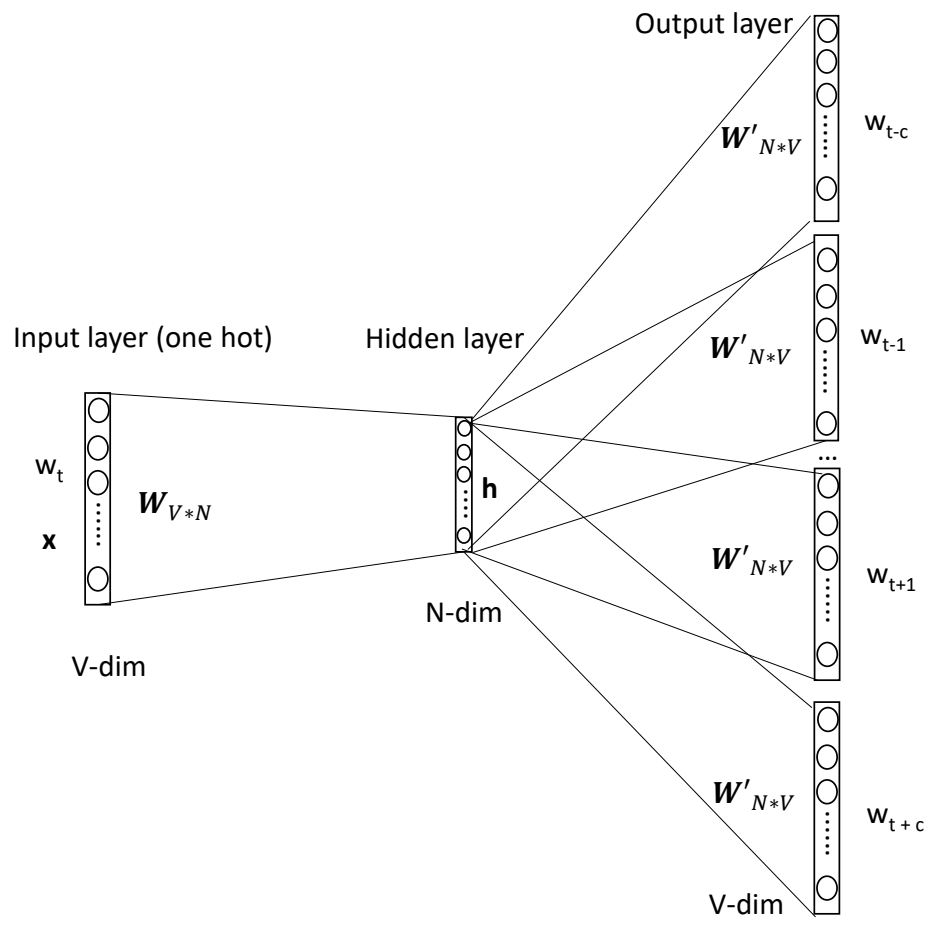

**Figure 1. The skip-gram model architecture**

Where, for each student, $s$, the average loss is calculated over the input screens at each time slice, $t$. The loss for a single time slice of a student is the sum of the log of the model's probability of observing the screens within a time slice window size, $c$, of the current time slice. While the model is trained to minimize error in predicting the screens in context, the intended extract from the model after training is not its predictions of screens but rather the learned representations of the screens in the form of the weight vectors associated with each screen. These weight vectors, a product of the single hidden layer of the model, are the automatic featurization of the screen. Screens which have similar contexts become mapped to vectors of similar magnitude and direction in order to minimize the loss.

## CASE STUDY – *HABITABLE WORLDS*

*Habitable Worlds* is an introductory level 4-credit online lab science course developed by Dr. Ariel Anbar and Dr. Lev Horodyskyj. The course was developed from 2011-2015 in Smart Sparrow's Adaptive eLearning Platform (AeLP), a Powerpoint-esque development environment that gives instructors full control over content, layout, adaptivity, and learning pathways. The platform collects data while students interact with the educational content, and data are reported to the instructor via dashboard and through downloadable CSV sheets should the instructor desire to conduct more detailed analyses on a particular activity. Because of these capabilities, continual refinement of *Habitable Worlds* through data analyses, student feedback, and instructor intuition has been made possible. A full account of the pedagogy and design philosophy of the course can be found in Horodyskyj *et al.* (2018). The course reached a stable version by the Fall semester of 2015 with minimal subsequent changes. Despite improvements in course content through changes based on descriptive statistics like time-on-question, number of attempts, and the frequency of certain incorrect answers, general behavioural insights were difficult to come by due to the sheer amount of data that is available and the lack of algorithms for parsing the data into usable information.

Course content is divided into three types of activities: *Training*, *Application*, and *Project*. Most content in the course is linear in nature, with occasional adaptive pathways to remediate misconceptions or bypass mastered content. *Training* activities are activities that introduce students to new concepts and give them the opportunity to explore and experiment without penalty, usually through the use of surveys, short written text, images, short videos (<8 minutes), simulators, equations, problem sets, and virtual field trips (Figure 2). Points are accumulated as students pass milestones. Unlimited attempts are allowed; however, students cannot progress until they complete the required activity correctly. *Applications* are equivalent to quizzes, in that students are expected to demonstrate competency on them after learning in the associated training activities, with points are deducted for incorrect answers. The *Project*, essentially a final exam, is a comprehensive activity requiring students to locate rare habitable worlds in a field of 500 randomly generated stars. Course content is released on a weekly basis during the course's 7.5-week deployment period, and students have a week to complete newly released exercises. Training activities remain open and accessible for full credit for the entire term. *Applications* close off week by week and the project opens on the second week of the course and remains available the rest of the term (Figure 3).

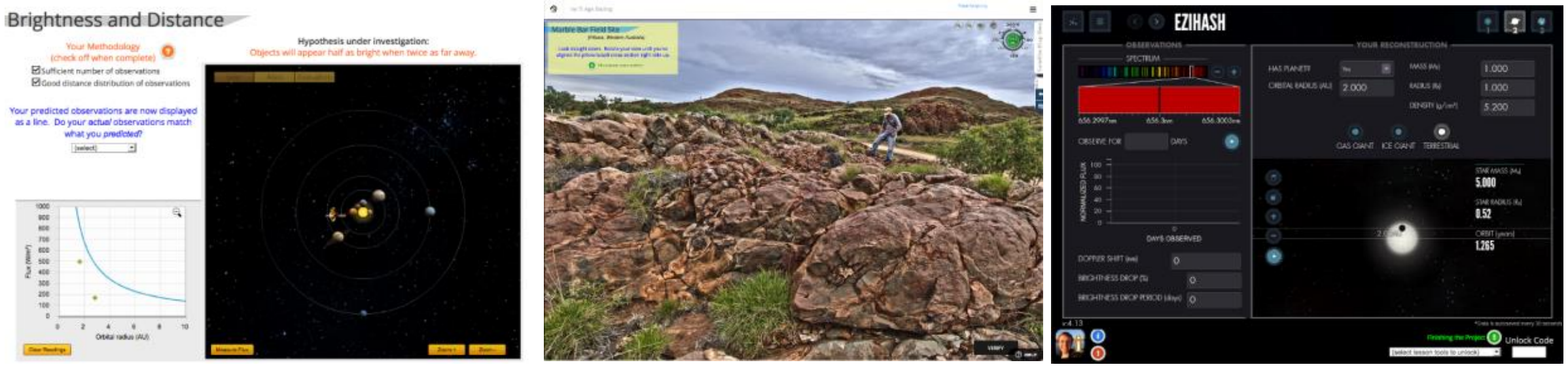

**Figure 2: (left) An experimental activity with a simulator involving a hypothesis, a check on the execution of their methodology, and an evaluation of their hypothesis. (center) An observational activity at a virtual field trip. Here, students are instructed to rotate their view to observe basaltic rocks in the field in a particular orientation. (right) The course project, where students are required to find rare habitable worlds in a field of 500 randomly generated stars.**

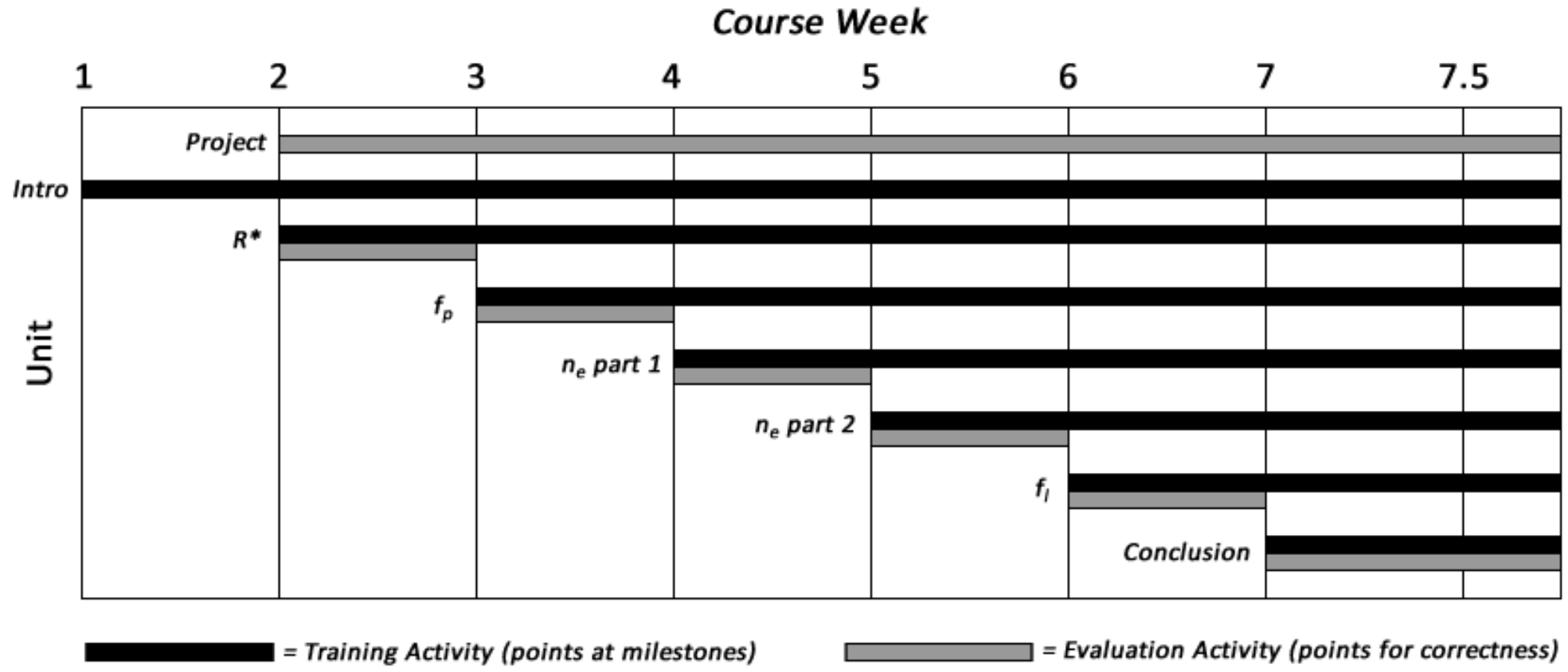

**Figure 3: Schedule of activities for *Habitable Worlds*. *Training* activities open week by week and remain open throughout the term. *Applications* open for one-week windows. The *Project* opens in week 2 and remains open throughout the term. Unit titles (R* = stars, $f_p$ = planets, $n_e$ = Earth-like planets, $f_l$ = Life) are based on the Drake Equation, a common astrobiology construct for exploring the question of life in the universe.**

The intended completion strategy for *Habitable Worlds* is to alternate between a training activity and its paired application, while accessing the project on a weekly basis as units are completed. However, students are not restricted to the prescribed path. Based on aggregate data generated by the AeLP, student behaviour observed on a paired discussion board, and general intuition, it was hypothesized that students who failed the course were taking non-optimal pathways; however, the nature of those pathways had been opaque to the course instructors to this point.

## Data

We used the time-stamped interactions (Table 1) of 778 anonymized students from two offerings of *Habitable Worlds,* Fall 2015 and Spring 2016. A skip-gram model[2] was used to learn continuous representations of course materials from 1.4 million temporal interactions of students with 1,644 pages of the course, called screens, within 67 different lessons. This model had several tuneable hyperparameters that changed the learned representations and thus the resultant visualizations. These were the context window sizes and the number of nodes in the hidden layer (vector size). Information on students' final grades and contributions to the discussion board were also included in the dataset.

| Field | Description |
|---|---|
| userID | Unique identifier for each student |

| userID | Example sequence of screenIDs |
|---|---|
| Sam | s:1 s:2 s:3 s:2 s:3 **s:4** s:5 s:6 s:7 … |

[2] Our research code utilizes the python genism implementation of word2vec

| | |
|---|---|
| screenID | Unique identifier for each "screen" in a lesson |
| interactionID | Unique chronologically ordered identifier for an event recorded by the AeLP (typically triggered when students click a "Check" button on the screen) |
| Erica | s:1 s:6 s:5 s:6 s:3 s:6 s:7 s:15 **s:14** s:15 s:12 s:15 s:12 s:15 s:16 s:21 … |
| Paulo | s:1 s:2 s:3 **s:4** s:5 s:6 s:7 s:8 s:9 s:10 s:11 s:12 s:13 **s:14** s:15 s:7 s:8 s:9 s:10 s:11: s:12 s:13 **s:14** s:15 s:15 s:15 s:16 … |

**Table 1: Fields used to construct the dataset (left). Example sequences of screenIDs (right). These sequences, specified one row per student, comprised the dataset used to train the skip-gram model. Application (graded) screens are in bold.**

## Parameter Tuning & Visualization Evaluation Results

There is no established rule of thumb as to the appropriate hyperparameters for skip-grams to use for bringing out visually salient patterns in a dataset. As a result, part of our methodology involved finding good values using a limited range of hyperparameters to create a set of 21 representations and respective t-SNE visualizations which were rated for their usefulness by the co-creator of the course (2nd author) on a five-point scale (Figure 4) in similar fashion to Géryk (2015).

As mentioned earlier, the guidance given to the expert rater was to favor visualizations that both confirmed existing intuitions and provided additional insight. After initially scanning all visuals to identify common patterns, three confirmatory patterns quickly emerged: (i) a grouping of week 2 quizzes – expected due to observations showing students racing to finish expiring activities before a hard deadline, which first happens in week 2; (ii) the beginnings of new units connecting to the endings of old units – expected based on observations of high-performing students beginning new activities as soon as they were released; and (iii) the splitting of a particular week 5 activity – expected, as the activity is excessively long. Visualizations were rated based on the presence of these three patterns. A visualization would receive extra points if, in addition to possessing the three patterns, the particular plot highlighted novel but explainable patterns.

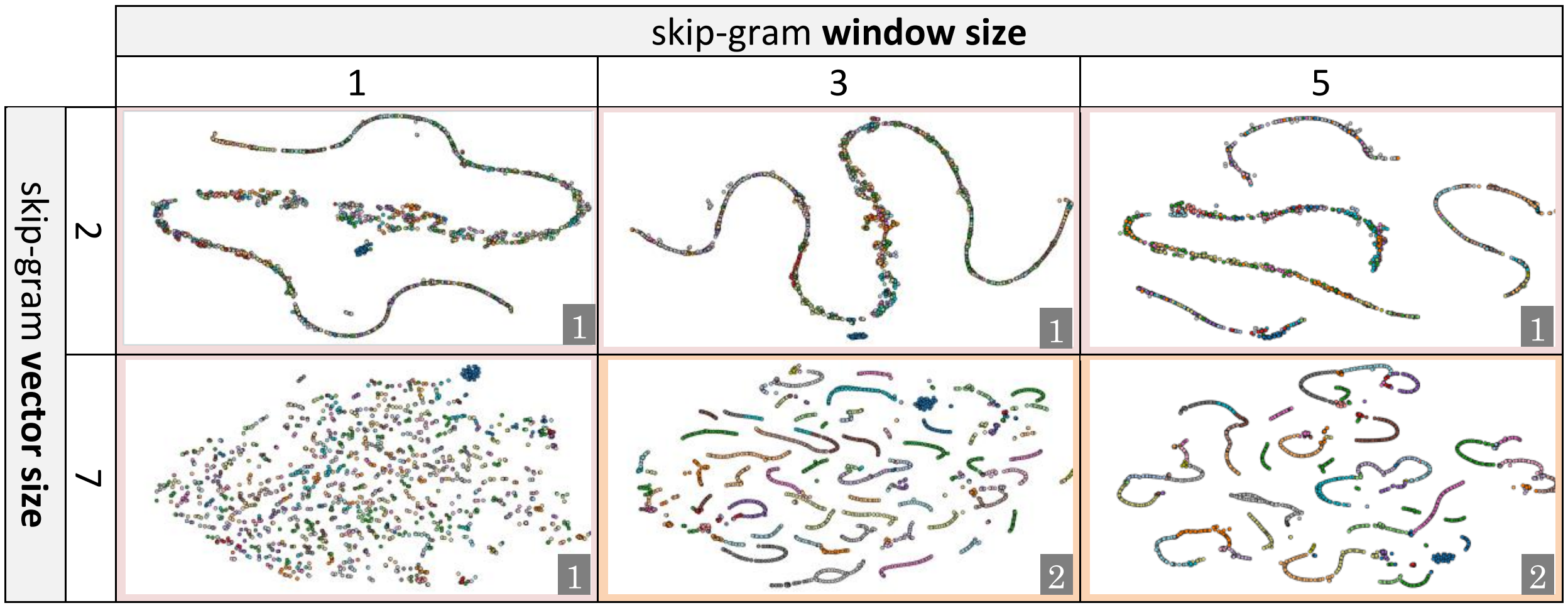

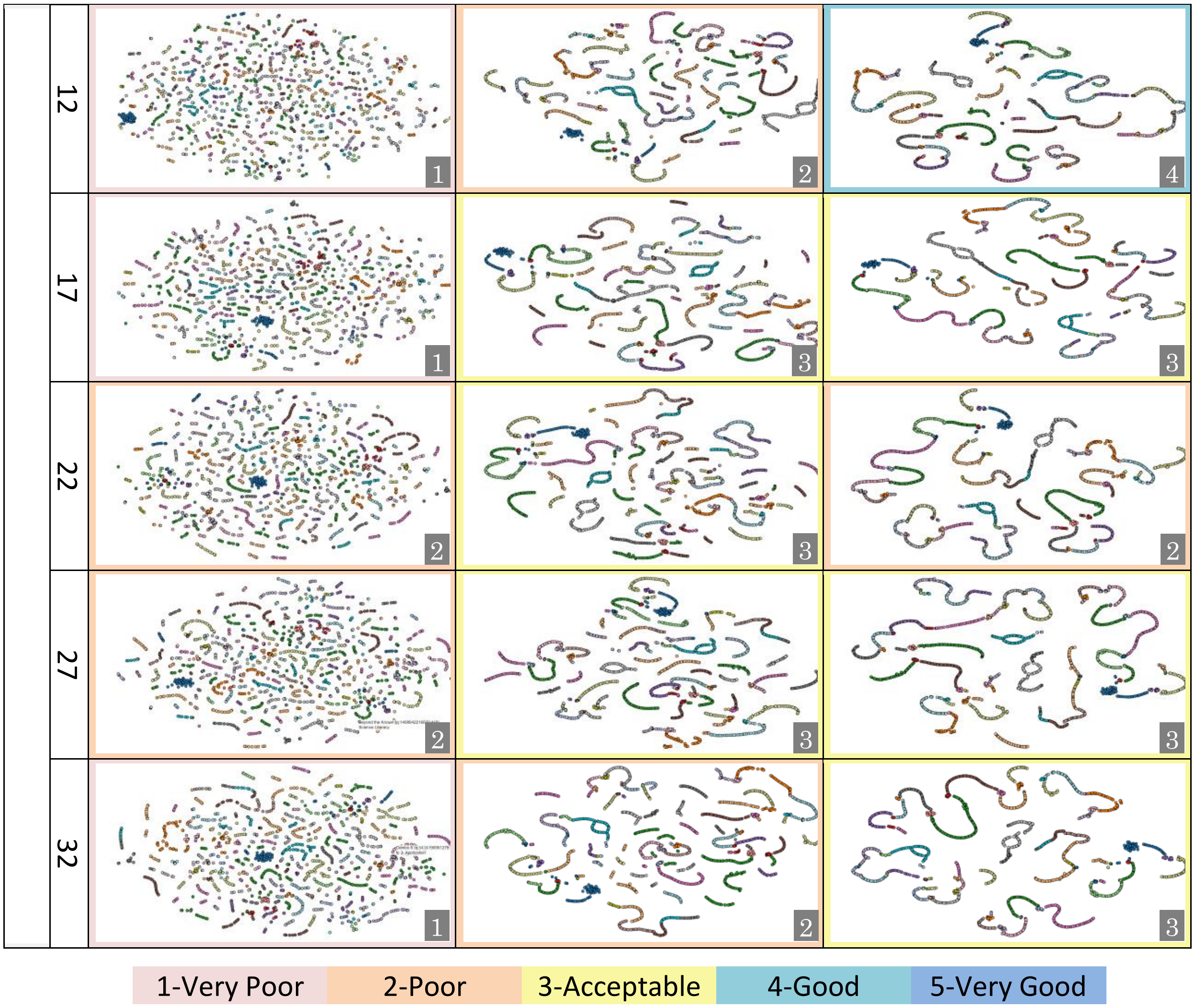

**Figure 4: Skip-gram t-SNE visualizations with a variety of window sizes (1-5) and vector sizes (2-32) with instructor usefulness ratings shown in the lower right corner. Each of the 21 scatter plots depict the complete set of elements in the course and their relationship to one another. Colours of plot points represent the lesson within the course to which each point belongs. Useful plots were those which depicted hypothesized and plausibly explainable relationships between course elements.**

Analyzing the results in Figure 4 from an algorithmic standpoint, the smaller vector size hyperparameters generally lead to simpler, more singular and linearly connected data points, representing the underlying intended sequencing of the course. There is an analogy here to principal component analysis. When only allowed to represent course elements with two continuous values, these values capture the most predominant structure, which is the courseware sequencing, followed by many students. Large vector sizes can bring out second- and third-order patterns of significance, which will be looked at more closely in the next section. When the vector size was two, window sizes of one and three produced more linear, connected plots than a window size of five. With a smaller window with which to learn representations, patterns consisting of many behaviours cannot be easily considered and the representation takes on a form more common when constructing transition plots based only on the frequency of transition from one element to the next, effectively a context window of 1. At higher vector sizes, a higher context window size struck a balance between providing a paucity of sequence patterns and retaining identifiable and anticipated patterns of behaviour.

## Results

In summary, the highest rated visualization was produced by a skip-gram with a window size of five and a vector size of 12. Other high vector sizes, paired with high window sizes, also yielded visualizations with useful information. Low vector sizes yielded excessively sequential plots, while low window sizes yielded plots with almost no connections between screens.

With a model of satisfactory representational quality in-hand, the following sections further interrogate this informational artifact to investigate more pointed questions regarding student behaviour and its connection to the course's pedagogy. Each iteration of the analysis adds an extra layer of complexity in order to answer research questions of increasing specificity.

## Iteration 1 – Inspecting the Highest Rated Visualization

The annotated visualization, most highly rated from the parameter tuning, is shown in Figure 5, with the prescribed course sequence marked with black arrows and the beginnings of new units starred. Essentially, the plot shows that students tend to approach a course that was designed to be linear in a mostly linear fashion. The visualization reveals behaviours that were previously intuited from aggregate data and discussion board observations. Although helpful in visualizing and confirming student behaviours that had been previously assumed, the skip-gram revealed only limited additional information for course improvements, mostly related to which lessons ought to be split due to being too lengthy.

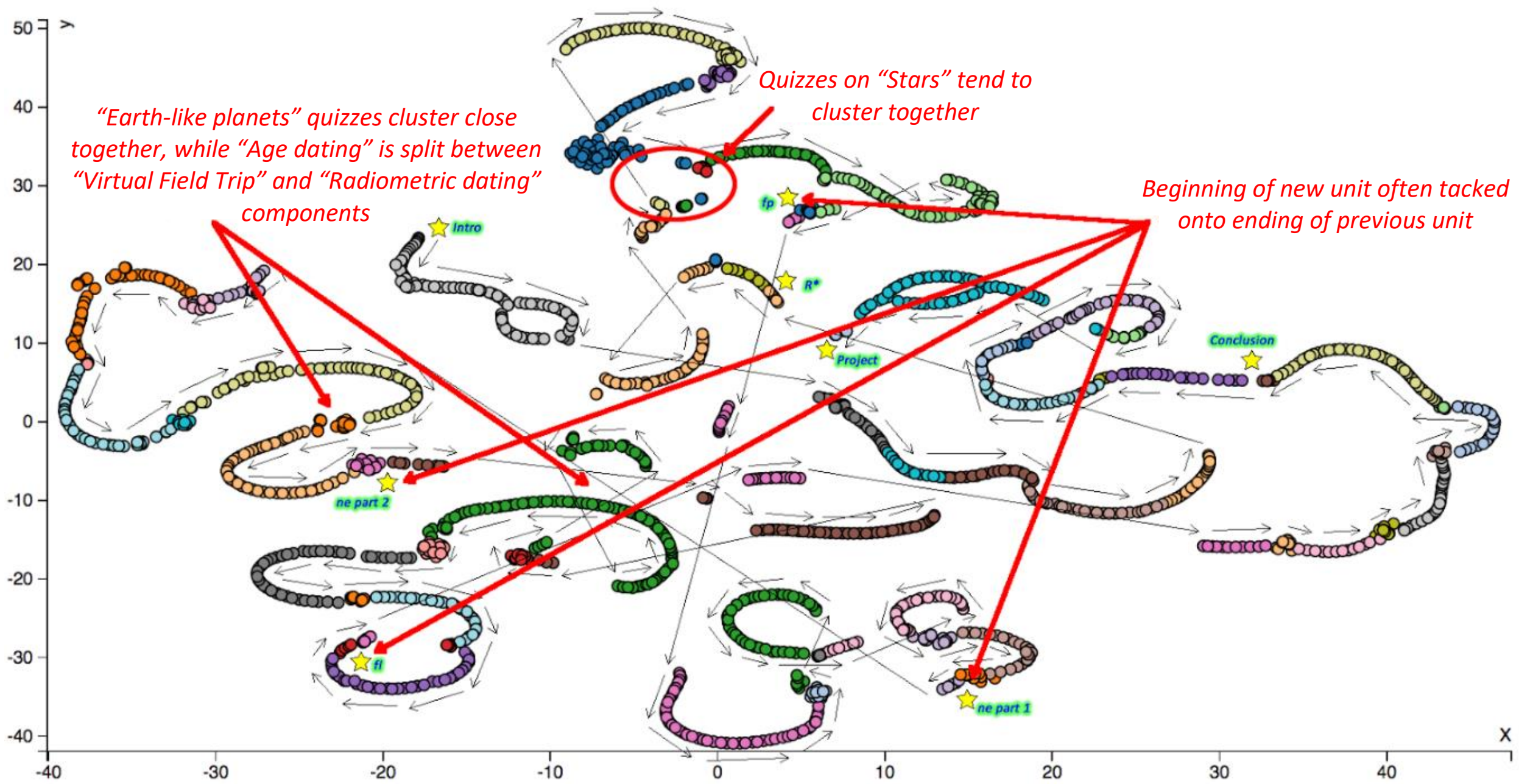

**Figure 5: Best rated visualization of course elements from parameter tuning stage with instructor annotations. Axes are abstract and represent dimensionality reduced vector values. Gray arrows indicate the sequence of content in the course, stars indicate the beginnings of a new unit, and red notes indicate patterns of interest. Each lesson is represented by a different colour. The typical pattern consists of a long linear sequence representing the training activity followed by a more clustered quiz activity. Course material progresses from *Intro* to *R** to $f_p$ to $n_e$ *(part 1)* to $n_e$ *(part 2)* to $f_l$ to *Conclusion*.**

**Iteration 2 – Differentiating Behaviours Exhibited by Passing and Failing Students**

Failure in courses is often attributed to student difficulties with content or disengagement. Instructors often have little insight into why a student has failed a course, aside from either observing that a student has stopped attending class or performed poorly on an assignment. Analytical data from learning management systems and MOOCs have provided new windows into student behavior, particularly when interacting with digital content. Previous informal analyses of *Habitable Worlds* offerings showed differences in engagement between passing, failing, and withdrawing students, in addition to level of content mastery and "attendance" based on whether course content was accessed. Passing students fully engaged with a majority of the content, while students who withdrew engaged with little, if any content. Failing students, surprisingly, showed persistence in the course, often engaging with content week after week, though not completing it successfully.

The second iteration of the skip-gram featured the incorporation of pass/no pass information, creating two of each course element. If a student had passed the course, each element in her sequence would be prefixed with a ‘p’ (e.g. “p-s:12”) or an ‘n’ if she had not passed. A single model was trained to represent these elements along with a single dimensionality reduction, but since no student sequence consisted of both pass and no pass elements, the two sets of representations do not substantively share the same space as they never appear in each other’s context windows. This single projection of passing and non-passing student course representations can be seen in Figure 6. Although content in *Habitable Worlds* is linear, individual training (10-60 screens), application (~5 screens), and project (~5 screens) lessons can be accessed in any order, if desired, during any given week. The intended course pathway is to complete a training activity followed by its associated application activity, repeated for each topic (between 4 and 6 for each week). Once the unit is complete, the associated component of the project can be completed. Release of units of the course is time-gated, but lessons within the units, once released, are not gated.

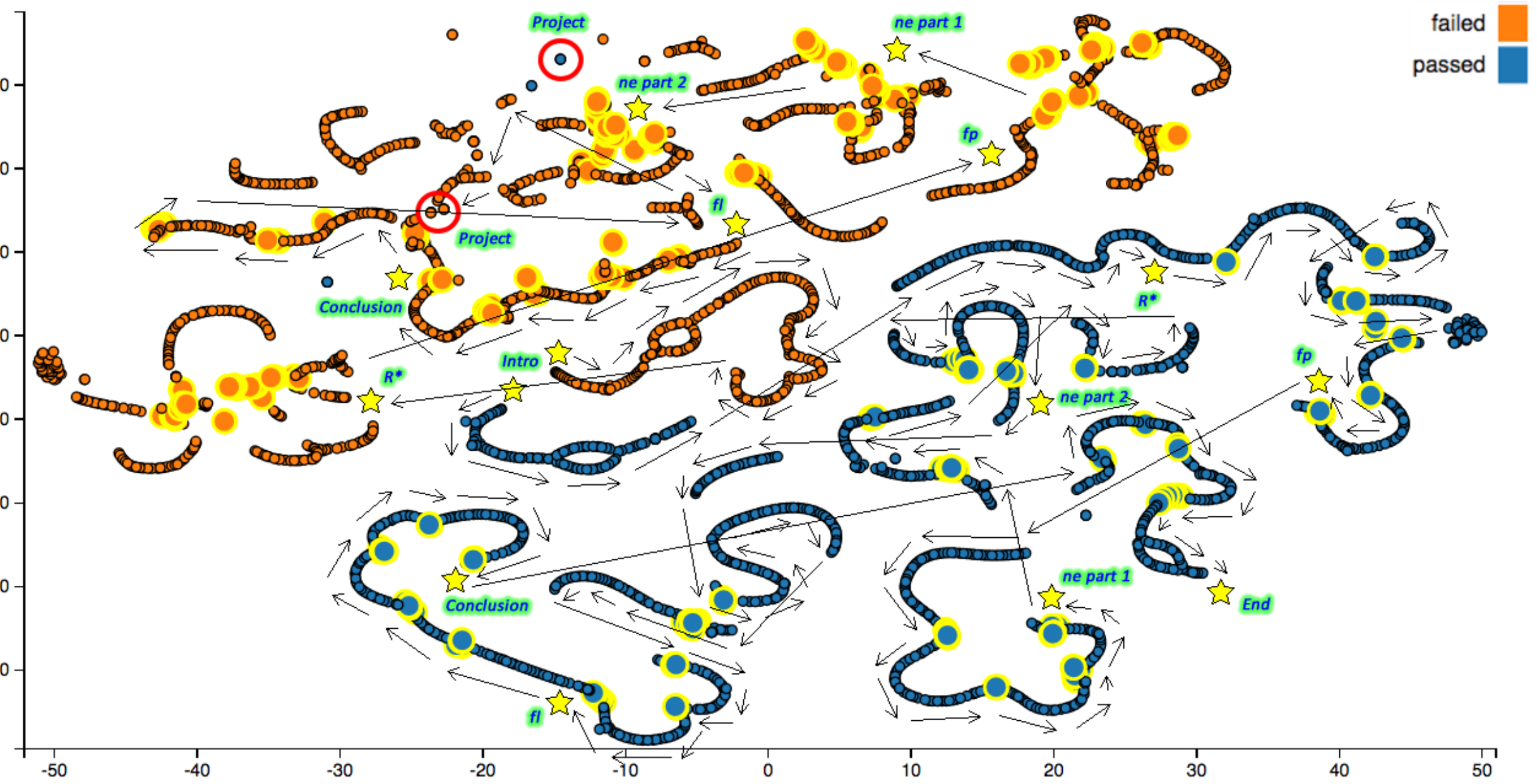

**Figure 6: Behaviour of failing (orange) and passing (blue) students. Screens associated with training activities are outlined in**

**black, quiz/application screens are highlighted in bright yellow, and project screens are circled in red. Passing students show a linear progression through content, with quizzes/applications appearing sequentially after their associated training activities. Failing students show a hub-and-spoke pattern for most units, with training activities converging onto a cluster of quiz/application activities for that unit. The project clusters far from activities for passing students (indicating random access during the semester), and clusters close to the concluding unit for failing students.**

The introductory unit of the course, completed in week 1, does not have any associated applications, only training activities. Paired applications begin with week 2 content. For both passing and failing students, week 2 (R*) applications cluster together, indicating that students are accessing them closely together in time. This may indicate that students are racing the deadline and attempting to complete the hard-deadline content (training activities have no deadline). Passing students seem to realize that this is not an optimal strategy and switch to a more sequential approach in subsequent units, where they complete training activities followed by applications as they appear in the sequence of the unit, a pattern consistent with self-regulated learning. Failing students, however, make this switch much later in the term. This results in a hub-and-spoke pattern for each unit on the visualization, where applications cluster together, with training activities radiating off of them. This indicates that students may be attempting to complete the applications first, only proceeding to the training activities in order to find the answers to the applications.

In addition, there are differences between passing and failing students in the visualization with respect to the final project. The project is released in week 2 and is a fairly complex endeavour, requiring students to utilize skills learned in almost every unit of the course and assembling those skills into a methodology to find a handful of habitable worlds in a field of hundreds of stars. The optimal strategy is to engage with the project early and complete components of the project as the concepts are learned. For passing students, the project clusters quite far away from the rest of the course, indicating that they are accessing the project throughout the course, hence the lack of association with any particular week's activities. For failing students, however, the project clusters very close to the Conclusion unit, indicating that failing students do not engage with the project until all other course material has been completed.

Overall, this visualization, where students are differentiated based on their grade, revealed that although both groups of students take the precarious strategy early in the course, passing students subsequently adopt a more optimal strategy while failing students do not, continuing to struggle week after week. These results are consistent with previous work showing that successful students in active learning settings utilize deep learning strategies while struggling students in the same context suffer because they are adopting shallow learning strategies (Gašević, Jovanović, Pardo, & Dawson 2017). This is, however, inconsistent with previous work indicating that students move towards less effective learning strategies over time (Jovanović, Gašević, Dawson, Pardo, & Mirriahi 2017). This may be a function of course design, where shallow learning strategies are consistently frustrated by dead-end alternative pathway and misconception loops, resulting in a course where the “path of least resistance” requires the continuous application of deep learning strategies.

### Iteration 3 – Discussion Forums

*Habitable Worlds* is paired with a discussion help forum on the Piazza platform. Students who are stuck or require a more detailed explanation on any course concept can post their questions to the forum, where either an instructor or a fellow student can reply. Average response time is 5-10 minutes during most of the day. Instructors typically offer assistance for training activities only, while fellow students are allowed to offer assistance on training, applications, and the project. Because most of the evaluations are generated randomly using equations and algorithms in the AeLP, students can exchange techniques for solving problems, but not answers as no two students have the same problem sets. This results in a collaborative environment in which students and instructors work together to learn and master skills and concepts, mimicking a real scientific environment.

Forum posts, but not views, were also part of our time-stamped data and were added to the skip-gram input sequences to understand how passing and failing students utilized the discussion board (Figure 7). These representations were also learned with a single model but, for this analysis, the projection (i.e., dimensionality reduction) for each pass/no pass group was generated separately. With few exceptions, passing students' forum postings do not tend to cluster with any particular training/application combination. This suggests that passing students are posting on the discussion board only when they need help, and this differs from student to student, resulting in clustering of posts away from any particular lessons.

Failing students' forum postings tend to have a slightly stronger association with applications than for passing students, although this is not universal. This indicates that when running into problems, failing students reach out for assistance, a desired behaviour. However, because this tends to happen during the *application* activities rather than the training activities, this indicates that these students have likely not mastered the content in the training activities, and so when asked to apply their knowledge, they struggle. The reason for the lack of content mastery is unclear. Piazza records the number of posts and threads "read" by students but does not timestamp this particular activity. Many students have self-reported that they do not post on the board often because they start activities late and find that when they become stuck, their questions have already been answered on the forum days earlier. Training activities cannot always be randomized because they are teaching conceptual frameworks, and these frameworks often have singular explanations (e.g., there are only so many ways to teach "higher temperatures = more evaporation"). If a student is using the discussion board as a crutch to complete difficult training content by doing what previous students have reported as resulting in success without understanding *why* it led to success, it is likely that they are not internalizing the concepts of the course. Hence, when they reach the application, which creates randomized activities based on the underlying concepts, it is likely that they will continue to struggle in spite of reaching out for help because they never gained a deeper understanding of the underlying concept that is being tested in the application.

**failed**

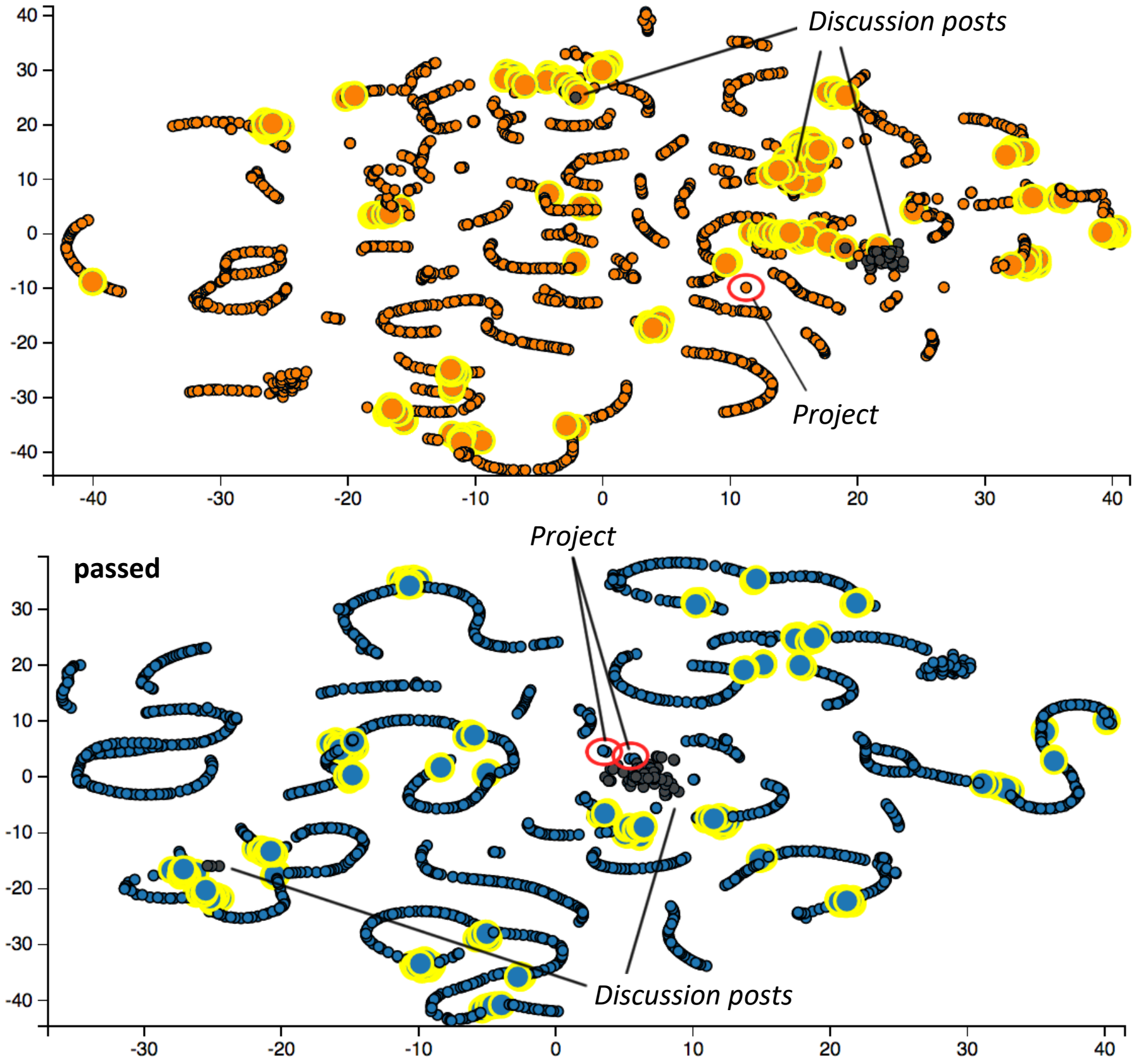

**Figure 7. Behaviour of failing (orange) and passing (blue) students. Screens associated with training activities are outlined in black, quiz/application screens are highlighted in bright yellow, project screens are circled in red, and forum discussion posts are grey-filled circles. (top) Failed student forum behaviour, showing some correlations between forum posts and application activities, and little association between forum posts and project work. (bottom) Passed student forum behaviour showing little association between forum activity and application activities, but strong association between forum posts and project work.**

There is a notable reversal of this behaviour for the course project. The project is a summative assessment, equivalent to a final exam. Students are presented with 500 randomized stars and are required to find rare habitable worlds, synthesizing techniques learned in the class, and reinforced in applications, into a search strategy. Students can develop and exchange strategies and check each others' work, but they cannot trade answers because each star set is unique to each student. A collaborative environment is not necessary to successfully complete the project and is purely optional. For passing students, their project pages cluster very closely with forum posts, indicating that these students are taking advantage of the collaborative environment available to them. This matches observations where instructors have seen collaborative groups forming and developing strategies together. For failing students, project pages do not show a close correlation with discussion posts. This clustering pattern of project pages suggests that failing students are *not* collaborating on an activity for which collaboration is most beneficial. They may be skimming information from other discussions for strategies, but do not seem to be engaged in the process of synthesizing strategies themselves.

## *Habitable Worlds* Case Study Discussion and Conclusions

The visualization methodology provided corroboration of a variety of existing intuitions held by the instructor from interacting with students via discussion boards, e-mail, and in-person over the years that *Habitable Worlds* has run. The pathway visualization supported the expectation that students followed a mostly linear path through the course. The relative length of lesson paths, on a cursory level, helped to identify content that could be split into smaller pieces to benefit learners. More significantly, the analyses revealed substantial navigational differences between passing and failing students. The interpretation of differences depicted by the visualization was that passing students take the quizzes in tandem with the practice material while the failing students go directly to the quizzes and then seek out the answers in the learning material. This information can prove essential to formative course development.

*Habitable Worlds* was designed to be approached in whatever way a student finds most comfortable. A student can pace their work across an entire week or complete it all in a short burst. Students can work alone or together, either online or in-person. The course project can be completed throughout the entire semester or all at the very end. This design was intentional to allow maximum flexibility for students, many of whom are non-traditional and have significantly more responsibilities than a typical on-campus student. However, for many failing students, this design may be detrimental as it depends on a certain amount of self-discipline and awareness of personal limitations, a skill that is often not well-developed in novice learners. The representation learning visualizations confirmed an existing assumption about failing students' unsophisticated strategy towards engaging with the course project, but also revealed that this strategy was applied to the rest of the course content and likely the usage of the discussion board as well. Students who fall into this category may require a more structured approach or alternative incentivization in order to adopt successful study strategies for completing the course. Future versions of the course will focus on building better supports, such as an early module on successful course strategies, and course structure enhancements, such as adding flexible deadlines and enforced or incentivized content ordering. These planned modifications are expected to better support those who are not just struggling with the subject matter of the course but with learning to learn in an online environment.

## CONTRIBUTIONS AND DISCUSSION

We have introduced a methodology around the nascent field of representation learning visualization and its interpretation and applied it to the domain general topic of summarizing the temporal behaviours of learners. The intention of this approach was to require as few assumptions about behaviour as possible, instead allowing prominent features of learner navigations to surface organically with the aid of careful model and visualization tuning that took place between the researcher and the practitioner closest to the domain. By allowing a model to take the place of an expert in creating the abstract featurization of behaviour, novel insights could be made by the expert in his interpretation of the visualization.

This approach can be seen as a type of Human-Computer Interaction (HCI); or more specifically, a type of Human-AI interaction where (1) an AI learns a set of representations of the world (via skip-gram) that allow it to make predictions of behaviour, (2) a human expert builds confidence in the AI's epistemic validity by selecting the representation that exhibits information known by the expert (i.e., model selection), and (3) observations are made by the expert from the representation visualization that support novel insights about behavioural patterns. Like any other expert, an AI can be wrong, and corroborating support for an interpretation should be gathered before a pattern is considered a bona fide generalizable phenomenon. Nevertheless, a machine learned representation can serve as a unique perspective on behaviour that can prove to be an indispensable companion source of information in design.

## FUTURE WORK

The machine learned representations in this work, with the addition of assessment constructs, might be comparable to an epistemic frame (Shaffer, 2006) or a network relating constructs to one another. Future work could investigate the nature of cognitive assessment information encoded by a learned representation versus a more explicit expert representation (Shaffer & Ruis, 2017). Although screen sequences and discussion posts were utilized in this study, there is a multitude of additional data that can be used in future iterations of the work applied to courseware or other contexts in which temporal learner data are collected. These additional data include problem answer text and correctness information, granular activity involving within-page interactive widgets (e.g., simulations), and interactions and communications with peers and instructional staff outside of the discussion board.

An obstacle to adoption of our method by a broader audience of instructors, without researcher support, is the rather careful process by which the parameters of the visualization were initially tuned. Since this was a collaborative research endeavour, guided by the educational theory informed goals of the instructor (Hillaire, Rappolt-Schlichtmann, & Ducharme, 2016), it was justified for this laborious effort to be undertaken by a member of the instructional staff; however, this should not be a pre-requisite, in practice, for a typical instructor to take on before interfacing with these analytics. The technique needs to be applied to more courses and more settings that vary by the type of students, course material, and platform to ascertain if common useful settings emerge in general contexts which would reduce or eliminate the upfront manual tuning effort. It is also an open question whether a threshold level of analytical data fluency or light professional development is required in order for a domain expert to independently begin to make sense of these visualizations.

## ACKNOWLEDGEMENTS

This research was supported by the Bill & Melinda Gates Foundation. We thank Smart Sparrow and the Inspark Teaching Network for facilitating this work and providing anonymized student data. Development of *Habitable Worlds* was supported by the National Science Foundation (Award #1225741). This research was conducted in accordance with Arizona State University institutional review board (Study #3679).